\documentclass[%
 reprint,
 amsmath,amssymb,
 aps,prl,
]{revtex4-2}

\usepackage{graphicx}
\usepackage{dcolumn}
\usepackage{bm}
 \usepackage{tikz}
 \usepackage{subfigure}
 \usepackage{adjustbox}

\usepackage{glossaries}
\newacronym{md}{MD}{molecular dynamics}
\newacronym{lj}{LJ}{Lennard-Jones}
\newacronym{mct}{MCT}{mode-coupling theory of the glass transition}

\begin{document}

\title{Shear-Rate Dependent Surface Tension of Glass-Forming Fluids}

\author{Linnea Heitmeier}
\author{Thomas Voigtmann}%
 \email{thomas.voigtmann@dlr.de}
\affiliation{%
Institut für Materialphysik im Weltraum, Deutsches Zentrum für Luft- und Raumfahrt (DLR), 51170 Köln, Germany
}%

\affiliation{%
Department of Physics, Heinrich-Heine Universität Düsseldorf, Universitätsstraße 1, 40225 Düsseldorf, Germany}%

\date{\today}

\begin{abstract}
We investigate the interface of a glass-forming fluid showing
non-Newtonian rheology.
By applying shear flow in the interface, we detect that the surface tension
depends on the shear rate.
Importantly, the standard way of determining surface tension from
the pressure drop across the interface gives rise to an
effective surface tension in the non-Newtonian fluid that mixes
bulk and interface properties.
We show how the pressure anisotropy can be used to clearly define the
bulk and interface regions and extract a genuine shear-rate dependent
surface tension. The results have implications for measurement techniques
related to interfacial rheology of complex fluids.
\end{abstract}

\maketitle



The surface of glass-forming fluids is home to many
intriguing dynamical phenomena.
Glasses are covered by layers of enhanced molecular mobility \cite{Tian.2022,Zhang.2018}, providing an amorphous analog to surface melting,
the physical mechanism conjectured to facilitate ice
skating at low temperatures.
The highly mobile layers near the surface enable the fabrication by
layerwise deposition of
ultrastable glasses with unique mechanical properties \cite{Swallen.2007,Berthier.2017,RodriguezTinoco.2022},
and are a major factor in determining the properties of polymer films \cite{Ediger.2014}.
The surface induces a dynamical penetration depth into the bulk that
changes non-monotonically with temperature and allows to disentangle
the change in mechanisms of relaxation close to a dynamical cross-over
temperature $T_c$ \cite{peng2022nonmonotonic}.

Surface tension is a key parameter characterizing interfaces of
complex fluids \cite{Fuller.2012}, but not many theoretical
studies address it in the dynamical cross-over regime
close to the kinetic arrest transition.
This is despite the significance in applications for films and coatings,
in interfacial rheology in general \cite{Jaensson.2021},
for certain 3d-printing techniques \cite{Ragelle.2018}, and also for
theoretical concerns:
The surface tension of amorphous structures in contact
with each other is posited by some theories to play a key role in the
dynamical cross-over from supercooled liquid to glass \cite{Cammarota.2009,Ganapathi.2018}.

Glass formers typically are shear-thinning and yield-stress fluids, i.e.,
their viscosity strongly decreases with the flow rate, and they flow
only above a certain threshold stress close to $T_c$
as the structural relaxation rate of the quiescent fluid drops below
the imposed flow rate
\cite{Voigtmann.2014,WagnerBook2}.
Reliable experimental data for the surface tension of such non-Newtonian
fluids are rare, since
the emerging yield stress impedes measurements \cite{Joergensen.2015}.
Due to the associated slow relaxation time scale of the fluid, typical
measurements \cite{Boujlel.2013,Joergensen.2015} are prone
to hysteresis effects. For example, J{\o}rgensen \textit{et~al.}
\cite{Joergensen.2015} used a liquid-bridge tensionmeter and found
the apparent surface tension of carbopol dispersions to be
systematically higher in expansion than in compression. They rationalized
this finding with an elastoplastic model of the fluid, and attributed
it to the existence of a yield stress.

This poses a number of questions: first, how does the surface tension,
a parameter that is typically considered given by the ``static''
pressure difference across the interface (see below), couple to the
non-Newtonian rheology of highly viscoelastic fluids, a typical
``dynamic'' effect in the bulk? Second, since the bulk rheology of
the shear-thinning fluid depends very sensitively on the shear rate,
what is the effect of fluid flow on the (apparent) surface tension?

We address these questions by \gls{md} simulations of a prototypical model
of a glass-forming fluid (involving no polymeric or suspension effects),
in a simple setup involving a planar surface.
Our simulations reveal how an apparent surface tension arises in
non-Newtonian fluids that is a mixture of bulk rheology and genuine surface
effects.


\begin{figure}
\includegraphics[width = 0.9 \linewidth]{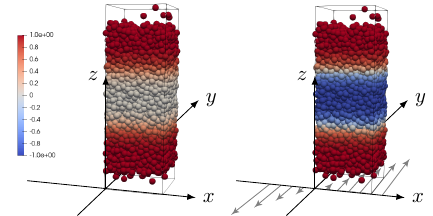}
    \caption{Snapshots of the simulation setup without (left) and with (right)
    imposed shear. Only the central part of the box along $z$ is shown for
    clarity. We impose 'in-plane' shear, as shown by the arrows in the
    coordinate system.
    Particles are colored according to $\Delta p=p_z-(p_x+p_y)/2$ normalized
    to the interval $[-1,1]$, see Fig.~\protect\ref{fig:normalstress}.
    \label{fig:snapshot}
    }
\end{figure}

We consider the standard Kob-Andersen binary \gls{lj} mixture
\cite{kob1995testing} at fixed number density $\rho=1.2$
using the open-source package LAMMPS \cite{plimpton1995fast}.
Units of length, $\sigma$, time, $\tau_0$, and energy are all in standard
\gls{lj} units related to the larger particles.
The simulations start from bulk liquids in the $NVT$ ensemble, that are
equilibrated at $T_i=2.0$ for at least $500\tau_0$,
cooled to the target temperature $T$, and then equilibrated again for
up to $10^6\tau_0$ (depending on $T$).
We use $N=5000$ ($N=10000$) particles corresponding to a cubic box
of size $L=L_x=L_y=13\sigma$ and $L_z=24.6\sigma$ ($49.2\sigma$) for the
small (large) system.
For the determination of the surface tension from capillary theory,
alongside the small system, a wide system with $L=18\sigma$
with $N=9564$ particles was used.

After equilibration, the simulation box was enlarged to $L_z=160\sigma$
keeping the fluid in the center of the box where around $z=0$,
bulk properties are recovered \cite{peng2022nonmonotonic}. The interfaces
were then relaxed for $250\tau_0$, before measurements were performed.
Shear flow with rate $\dot\gamma$
is imposed in the $(x,y)$-plane tangential to the surface
(see Fig.~\ref{fig:snapshot}) using the SLLOD equations;
this is also motivated by the tangential character of the
surface tension \cite{navascues1979liquid, birdi1997surface, gennes2004capillarity, durand2021mechanical, Marchand.2011}.
Note that our shear protocol keeps the total surface area constant, which
is conceptually important for viscoelastic fluids \cite{Marchand.2011}.
Surface tension values were averaged over at least $2000\tau_0$.
Inspection of the $z$-dependent density profiles revealed no
particle-species segregation.

We use two independent methods to determine the surface tension
$\hat\sigma$:
first from the pressure difference as suggested by
equilibrium statistical physics \cite{RowlinsonWidom},
\begin{equation}\label{eq:pressuredifference}
    \hat \sigma = \frac{1}{2} \int dz\,\left[ p_z(z) - \frac{1}{2} \left(p_x(z) + p_y(z)\right)\right]\,,
\end{equation}
where the integral extends from the bulk of one phase (the vapor) to the
other (the fluid). A factor of $1/2$ is included due to the second interface
in our simulation setup.
Note that this integral assumes the
integrand to vanish in the bulk regions, which is the case in the
isotropic Newtonian fluid. Essentially we argue that this integral has to be
taken with care in the case of a non-Newtonian fluid with shear.

To emphasize this point, we also determine the surface tension via the
width of the density profile $\Delta$ \cite{evans1979nature, sides1999capillary, vink2005capillary}:
\begin{subequations}\label{eq:capillary}
\begin{equation}\label{eq:delta}
    \Delta^2 = \Delta_0^2 + \frac{k_B T}{ 2 \pi \cdot \hat \sigma}\cdot \ln(L/B_0)\,,
\end{equation}
where we fit the density profile according to
\begin{equation}
     \rho(z) = \frac{\rho_l}{2} -\frac{\rho_l}{2} \cdot \tanh \left( \frac{2 \cdot(z-z_0)}{\Delta} \right)\,,
\end{equation}
with the bulk liquid density $\rho_l$, and $z_0$ a fit parameter
for the position of the interface.
In Eq.~\eqref{eq:delta}, $\Delta_0$ and $B_0$ are unknown prefactors
determined from fits for two different $L$ (small and wide system).
\end{subequations}

We checked that our results recover the surface-tension
values for the one-component quiescent \gls{lj} fluid reported in
Ref.~\cite{lutsko2020classical}.
For reference, the bulk viscocity was determined using the standard Green-Kubo
relation \cite{evans2007statistical} in simulations of the bulk fluid,
using a correlation time of up to $100\tau_0$.

\begin{figure}
\includegraphics[width=\linewidth]{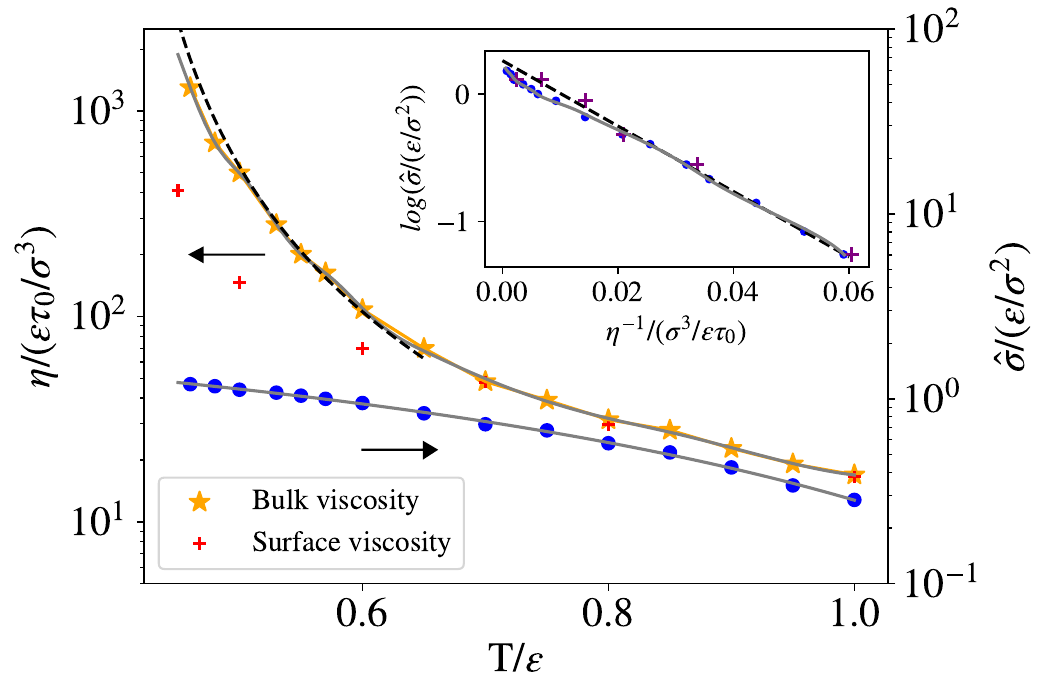}
\caption{\label{fig:viscosity}
  Bulk viscosity (star symbols) and surface tension (circles) of the
  quiescent system as function of inverse temperature.
  Solid lines are guides to the eye. A dashed line indicates the
  power-law predicted by mode-coupling theory,
  and crosses indicate the surface viscosity (see text).
  Inset: logarithm of the surface tension as a function of inverse viscosity
  (circles) and inverse surface viscosity (crosses).
}
\end{figure}

We begin by summarizing the temperature dependence in the quiescent system.
As the glass transition is approached, the viscosity of the bulk fluid
strongly increases; for around two orders of magnitude in the interval
$T=[0.5,1]$ studied here (stars in Fig.~\ref{fig:viscosity}).
The data is in the regime of the \gls{mct}: close to the cross-over
temperature $T_c$ of \gls{mct}, the viscosity shows power-law growth,
$\eta\sim|T-T_c|^{-\gamma}$ from which deviations would set in at lower
temperatures (a dashed line in Fig.~\ref{fig:viscosity} indicates the
power law with $T_c=0.4$ and $\gamma=2.35$).
In the same temperature interval, the surface tension (circles) increases by
almost a factor of $4$.
Notably, it initially increases alongside the viscosity in the regime of
temperatures $T\gtrsim0.5$, while at lower temperatures it decouples
and appears to saturate. Such saturation is in line with the
interpretation that the surface tension is a ``static'' fluid property
rather than a dynamical one coupled to the slow relaxation mechanisms
captured in \gls{mct}.

It has been stated that surface tension and viscosity can be related:
the empirical relation
$
    \ln(\hat \sigma) = \ln(A) + {B}/{\eta}
$
was proposed \cite{pelofsky1966surface,DiNicola.2018},
with coefficients $A$ and $B$ determined from fits.
This relation holds reasonably well in the not-too-viscous regime
($\eta\lesssim50$; inset of Fig.~\ref{fig:viscosity}).
For higher viscosities, we see deviations that we
attribute to the slow structural relaxation affecting the viscosity.

Particles close to the surface of a glass-forming fluid retain higher
mobility than in the bulk \cite{Tian.2022,Sun.2017,Zhang.2018}. It hence suggests
itself to define a ``surface viscosity'' that might bear a closer connection
to the surface tension since it will grow less strongly than
the bulk viscosity.

\begin{figure}
\includegraphics[width=\linewidth]{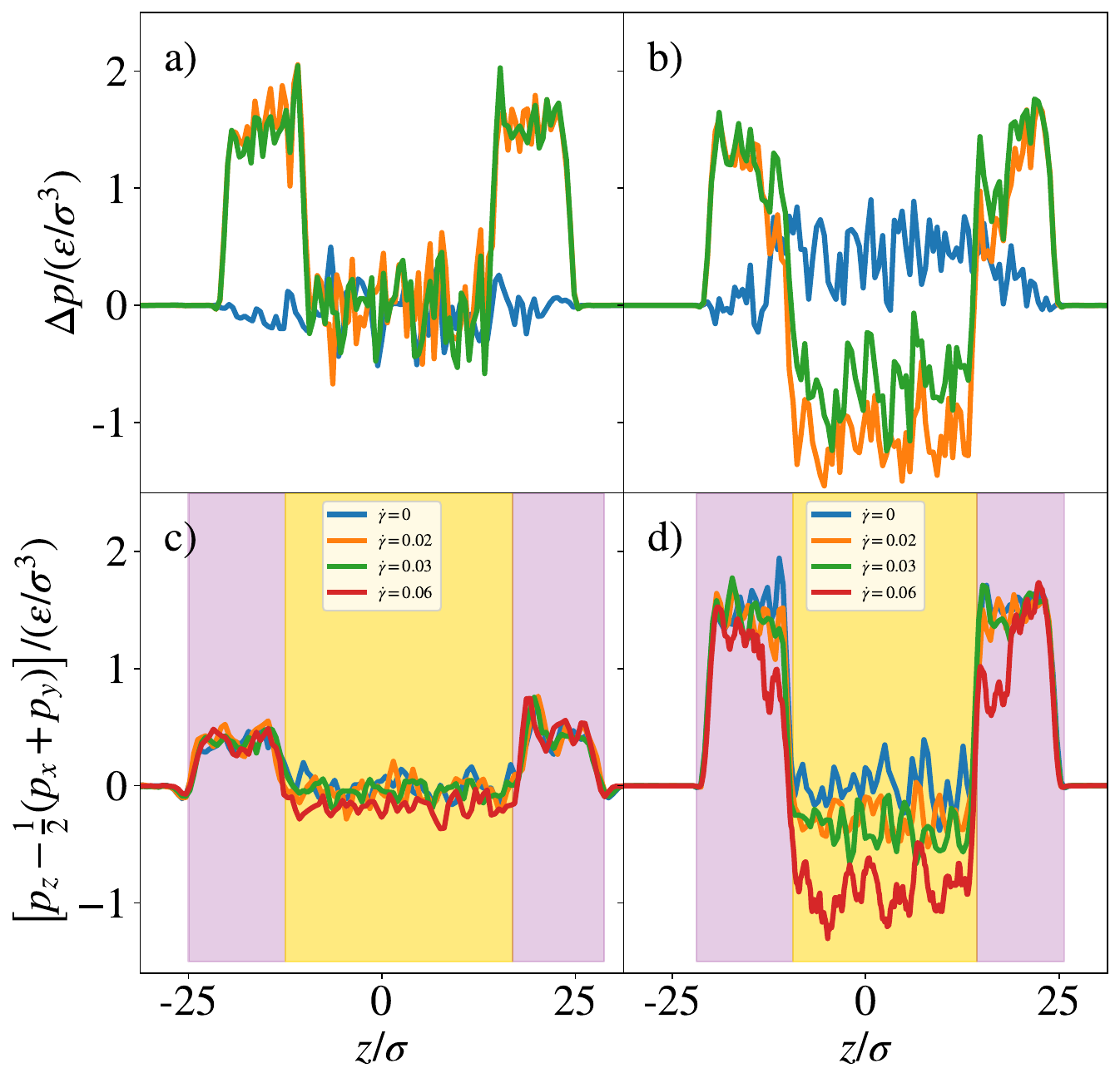}
\caption{\label{fig:normalstress}
  Pressure differences across the interface: Top panels show
  $\Delta p_{\alpha\beta}=p_\alpha-p_\beta$
  (blue: $\Delta p_{xy}$; green: $\Delta p_{xz}$; orange: $\Delta p_{yz}$)
  for temperature $T=0.6$ in (a) the quiescent system and (b)
  with shear rate $\dot\gamma=0.06$.
  Bottom panels show the relevant pressure difference for the surface
  tension, $-(\Delta p_{xz}+\Delta p_{yz})/2$, for
  (c) temperature $T=1$, and (d) $T=0.6$.
  Color shadings indicate the $z$-intervals identified as bulk and
  surface regions.
}
\end{figure}

We have defined a surface viscosity by constraining the Green-Kubo
integral to particles starting in $z$-layers close to the surface.
The notion of ``close'' can be made precise by looking at the
normal-stress differences, $\Delta p_{\alpha\beta}=p_\alpha-p_\beta$, where
$\alpha,\beta\in\{x,y,z\}$. Recall that in the isotropic bulk, all
$\Delta p_{\alpha\beta}=0$. Close to the surface, the values for $\beta=z$,
$\alpha\in\{x,y\}$, deviate from zero since the surface induces an
anisotropy. This allows to clearly distinguish a surface layer
(cf.\ Fig.~\ref{fig:normalstress}a), and hence unambiguously define a
surface viscosity.
This surface layer somewhat broadens upon decreasing temperature,
consistent with the existence of a growing length scale near the
glass transition \cite{peng2022nonmonotonic}.
We note in passing that the surface viscosity thus defined indeed follows the
empirical relation to the surface tension more closely (crosses in
Fig.~\ref{fig:viscosity}),
although approaching to $T_c$ a decoupling still appears to occur. We leave
this question for a future study.

Now we turn to surfaces with in-plane shear flow.
Glass formers are non-Newtonian fluids, featuring non-vanishing
normal-stress differences in the bulk.
Recall that we apply
a velocity in the $y$-direction, with gradient in the $x$-direction;
thus the first and second bulk normal stress differences are given by
\footnote{We use the ``tensile'' convention common in rheology, noting that
some authors use a ``compressive'' convention with the opposite sign.}
$N_1=\sigma_{yy}-\sigma_{xx}$ and $N_2=\sigma_{xx}-\sigma_{zz}$,
where $\sigma_{\alpha\alpha}=-p_\alpha$.
The Kob-Andersen system indeed shows $N_1=\Delta p_{xy}>0$ and
$N_2=-\Delta p_{xz}>0$ with
$|N_2|\approx N_1$ (Fig.~\ref{fig:normalstress}b, middle section).

Crucially, the surface-near region remains different: here, pressure
differences are related to the surface tension, and separate from the bulk.
The changes in these regions will be related to a change in surface tension
under shear. They become more pronounced at lower temperatures, as $T_c$
is approached, while at higher temperatures the effect vanishes
(lower panels of Fig.~\ref{fig:normalstress}).

However, the bulk values give non-trivial contributions to the surface-tension integral
that are specific to non-Newtonian fluids, and need to be taken into
account when determining the surface tension of the sheared fluid.
Also this effect becomes more pronounced when temperature is lowered:
At $T=1$, the fluid is nearly Newtonian, identified by vanishing normal-stress
differences in the bulk (Fig.~\ref{fig:normalstress}c); here we observe
that also the dependence of the pressure anisotropy in the surface region
on the shear rate is negligible. But at $T=0.6$, the non-Newtonian effects
are already pronounced: with increasing shear rates, increasing bulk
normal-stress differences are seen; also the surface regions begin to
show a shear rate dependence (Fig.~\ref{fig:normalstress}d).
In Eq.~\eqref{eq:pressuredifference}, hence an increasing bulk contribution
is present that leads to an apparent surface tension in the
non-Newtonian fluid.

\begin{figure}
\includegraphics[width = \linewidth]{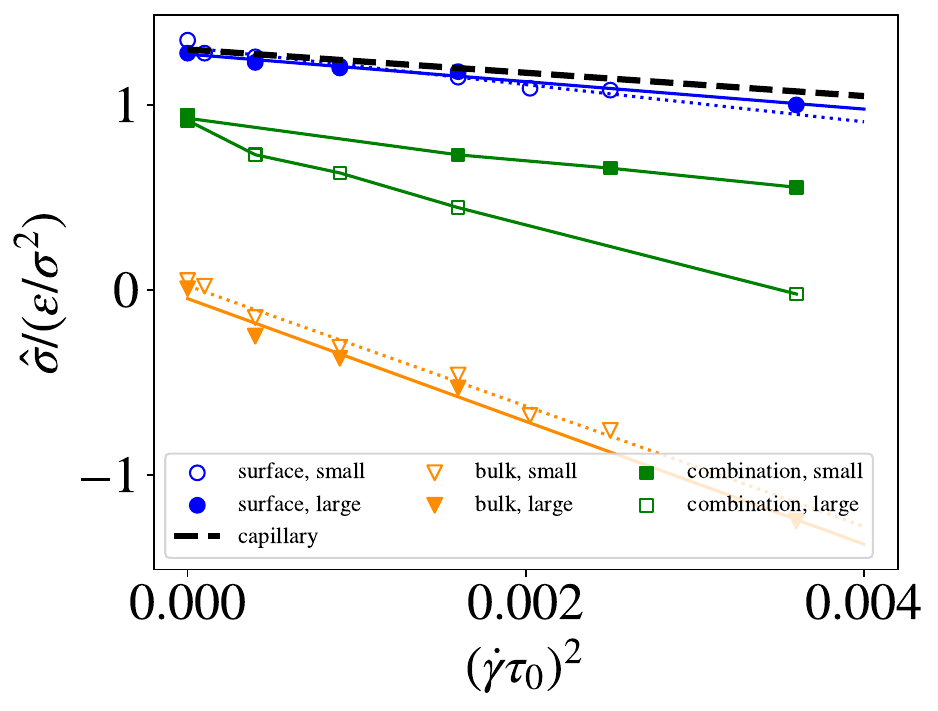}
\caption{\label{fig:sigma_gammadot}
  Shear-rate dependence of the surface tension, as a function of
  the square shear rate $\dot\gamma^2$, for the Kob-Andersen mixture
  at $T=0.6$.
  Blue symbols show values extracted from the pressure difference,
  a dashed line the trend extracted from capillary wave analysis.
  The contribution from the bulk of the non-Newtonian fluid (orange)
  leads to a system-size dependent effective surface tension (green), if evaluated by the pressure difference.
}
\end{figure}

The two regions in $z$ -- the surface layer identified by positive
$\Delta p_{\alpha z}$, and the bulk liquid where these quantities have
oppositve sign -- give rise to two different contributions to the
surface-tension integral (blue and orange symbols in
Fig.~\ref{fig:sigma_gammadot}).
Both depend quadratically on the shear rate, as is expected from the
symmetry of the problem under reversal of the flow direction.

Intuitively, it is the pressure drop across the surface layer that relates to
the surface properties. We confirm this by determining the surface tension
from an analysis of the density profile, Eq.~\eqref{eq:capillary}.
Since this procedure determines $\hat\sigma$ only up to a prefactor,
we have adjusted this to match the value obtained from
Eq.~\eqref{eq:pressuredifference} at zero shear rate.
The results for the $\dot\gamma$-dependence then are in very good agreement
with each other (dashed black line in Fig.~\ref{fig:sigma_gammadot}).

One notes that the bulk contribution to the pressure anisotropy is much
stronger in the non-Newtonian fluid than close to the surface; hence the
overall apparent surface tension in the thin film has a stronger
shear-rate dependence than the actual surface tension. It also demonstrates
a pronounced system-size effect
(green symbols in Fig.~\ref{fig:sigma_gammadot}). Larger systems
are more strongly influenced by the non-Newtonian bulk rheology, to the
point that the apparent surface tension might even vanish at large
shear rates.
It should be noted that this is not a destabilization of the interface;
it is rather a non-equilibrium signature of the driven system,
similar to what has been observed in active
fluids \cite{Marconi.2016,Speck.2016}.

Note that in our simulations we observe the surface tension to relax to
its stationary value after application of the shear flow very rapidly;
in particular, we observe this process to be much faster than the
structural relaxation time in the bulk. This further underlines that
the mobile surface layers decouple from the slow dynamics of the bulk,
but their shear-rate dependence still represents an intriguing coupling
between bulk and surface dynamics.


In summary, we have shown that the surface tension of a typical glass-forming
viscoelastic fluid shows a strong shear-rate dependence.
In non-Newtonian fluids, normal-stress differences in the bulk and
surface anisotropy of the stress tensor contribute
to an \emph{effective surface tension} that is measured using standard
techniques that are based on the pressure drop across a liquid film and
mix two different contributions.
One contribution is a genuine surface contribution, and it
comes from the pressure anisotropy in the layers near the surface.
In the simulation they can be clearly identified through the different
sign of $\Delta p_{\alpha z}$ with respect to the bulk.
In addition, there is a non-trivial bulk contribution arising from
non-vanishing normal-stress differences, a typical non-Newtonian fluid
effect.

This effect might explain why the determination of surface tension
values from pressure balances of bulk viscoelastic samples is difficult,
as in experiment it will be difficult to disentangle the pure surface
from a bulk contribution. Also, hysteresis effects as previously
reported might be arising from a non-Newtonian bulk contribution:
we have monitored the transient evolution of the pressure differences
after switching on the shear flow, and found no sign of slow relaxation
in the surface contribution. But the bulk quantities are known to exhibit
patterns of slow relaxation. Also, a changing volume-to-surface ratio,
as in techniques where a fluid droplet is expanded or compresses,
might see effects from different mixing of surface and bulk contributions.

Our findings should be relevant to various techniques
measuring interfacial rheological properties in films of complex fluids
\cite{Brooks.1999,Verwijlen.2011}
and even for the printability of bio-inks in surface-tension assisted
3d-printing techniques \cite{Naghieh.2021}.
For example, biofilms are known to exhibit complex rheology,
and characterization of their mechanical surface properties is an
important aspect in their growth and removal \cite{Geisel.2022,Charlton.2023}.
The complex interactions in such films can render the surface tension
anisotropic \cite{Jaensson.2021}, and our method of imposing in-plane
shear flow in different directions might be a straight-forward way to
interrogate the characteristics of the interface in such cases.

On a more theoretical note,
it has been shown that the surface of glass-forming fluids
reveals non-monotonic changes in the dynamical correlation length
governing glassy dynamics \cite{peng2022nonmonotonic,Tian.2022}.
However, this analysis was only based on local density profiles and
$z$-resolved density correlation functions. In principle, the
discussion of pressure differences as done in our work should give a
more natural handle on disentangling surface-induced from bulk effects
in such systems. Also, we leave for further discussion the fate of
the surface tension in the deeply supercooled regime (below the
\gls{mct} transition): eventually, the visco-elastic nature of the
glass-forming fluid will be felt, and as one approaches the soft solid state,
surface tension and surface energies need to be distinguished \cite{Style.2017}.
The study of the flow-rate dependence in the glassy regime can probe
the different time scales involved in the transition.

\begin{acknowledgments}
We acknowledge fruitful discussions with N.~J.~Wagner and
O.~D'Angelo.
The authors gratefully acknowledge the scientific support and HPC resources
provided by the German Aerospace Center (DLR). The HPC system CARO is
partially funded by ``Ministry of Science and Culture of Lower Saxony'' and
``Federal Ministry for Economic Affairs and Climate Action''.
\end{acknowledgments}

\bibliography{apssamp}

\end{document}